# The ADM version of GR at Sixty: a brief account for historians.

S. Deser
Walter Burke Institute for Theoretical Physics,
California Institute of Technology
Pasadena, CA 91125, USA
Physics Department, Brandeis University
Waltham, MA 02454, USA
deser@brandeis.edu

I review the meaning of General Relativity (GR), viewed as a dynamical field, rather than as geometry, as effected by the 1958-61"anti-geometrical" work of ADM. This very brief non-technical summary, is intended for historians.

I have been asked to update and simplify (no equations) a previous account [1] to render it more accessible for historians; there will necessarily be some overlap, but all equations are removed. Einstein's 1915 turning a force, gravity, into geometry was a unique revolution in physics: There were no more unmoveables: the last one, Space-time was not the rigid fixed *a priori* background for matter to move in, but a dynamical system interacting with its matter sources, but also (as against Mach) quite autonomous and capable of evolving in their absence—all this in a perfectly quantitative way. It is a curved manifold, governed by the Riemann curvature tensor, a quantity that expressed its degree of "bending" at each point; it consists of two algebraically distinct parts, the ("Machian") Ricci tensor governed by Einstein's field equations, and proportional to, hence determined by, the matter sources if any, while its other component, the Weyl tensor, was left undetermined, much as Maxwell equations do not fully determine the field strength from their currents. To be sure, the dimensionality of space-time remained unexplained, as it is to this very day; had it been D=3 rather than D=4, it would be purely Machian, as Ricci and Riemann curvatures coincide there! This almost mandates that we live in D>3, by the dynamical paradigm!

For a long time after the initial flurry of GR's famous explanation of then observable anomalies, the famous "three tests", let alone the built-in explanation of the identity of inertial and gravitational masses— an infinite set of predictions—the field remained fallow save for the occasional exact solution, besides the then little-understood Schwarzschild one. The reasons are interesting in themselves: there was no further observational incentive (cosmology and cosmic expansion were off to one side), but mostly because the impact of Quantum Mechanics on our understanding of the micro-world, so rich in observed and experimental data, soaked up all physicists. A telling anecdotal instance: on my 1953 arrival as a fresh PhD at the Institute for Advanced Study, its Director, Robert Oppenheimer cautioned me to have nothing to do with Einstein, its most famous denizen, nor with GR in any form, lest I become unsaleable on the job market! This admonition was easy to obey: Einstein was essentially invisible, and more important, one did not learn GR in Graduate School, most of whose faculty neither taught nor knew it (certainly at Harvard); little did Oppenheimer know that his greatest work—in the thirties with Snyder on Black Holes—would someday make his scientific name! Those were the glory days of QED

(Quantum Electrodynamics), QFT (Quantum Field Theory) and elementary particle (EP) physics in general. At the end of my stay there and after a second postdoctoral stint, at the Bohr Institute, however, there came a lull as QED wound down while EP's other sectors—the strong and weak interactions— were still in *statu nascendi.* There my GR tale begins.

In Spring 1958, I was completing a year's Instructorship at Harvard and looking for a new direction, as was my fellow graduate student and Schwinger advisee, the late Dick Arnowitt, then at Syracuse— coincidentally the home of one of the few US research groups in GR, led by Peter Bergmann; theirs was a dense forest of formalism that never really led to any tangible results, while Bryce deWitt was working solo on quantizing the theory, a very premature endeavor. Dick and I still knew nothing of GR, except that its weak field limit was the massless free spin 2 system in flat spacetime, right down our Schwingerian alley, so we decided to learn by doing and launched on a modern analysis of this model, first proposed by Fierz and Pauli in the thirties. We were of course guided by the Maxwell field, its spin one analog, and the prototype of a gauge field. It turned out that there were indeed many resemblances, but also that this very first higher (than one) spin theory was also profoundly different. In fairly short order we managed a complete dissection—it was a free field after all, that is a bunch of harmonic oscillators—two degrees of freedom (per space point) boiled down from the ten symmetric tensor components, as Maxwell's two were from its four vector ones, by gauge invariance in both cases. But just as the nonlinear Yang-Mills field is quite a bit more complex than the latter, so is full GR (infinitely) more complex than the former.

We wrote up our results, optimistically labelling the paper "1" of what would turn out to be more than a dozen. The rest of the series was made possible by a serendipitous invitation from John Wheeler in Princeton, who was also shifting his interests to GR, assisted by his brilliant student Charles Misner. Charlie heard our presentation and showed us some of his own results. At Wheeler's suggestion, we agreed to collaborate on future work, henceforth dubbed "ADM." Our joint endeavors primarily spanned some three years, 1958-61, and led to a steady flow of very different studies of full nonlinear GR as a field theory, rather than geometry. Of these, the most remembered and currently used aspects are the "3+1" decomposition of the metric tensor and its application to numerical calculations, the "ADM energy", at last defining the existence and properties of such a notion for —asymptotically flat— GR. This quantity had been sought in vain ever since Noether and Hilbert's attempts soon after GR's birth. [For a simple discussion of energy as arising by the interplay of Noether's two theorems, see [2].] However, the problem has ever been that relativists are not interested or versed in field theory, while conversely field theoreticians were not, until recently, interested in GR. Our work also first clearly established the (hitherto controversial) existence of gravitational radiation and its physical properties such as its wave zone etc, all in the strong field regime, the nature of the two degrees of freedom, the study of self-energies, the meaning of Hamiltonians in a generally covariant theories, and possible bases for the quantization of this unique system. I could go on summarizing the corpus, but instead refer the reader to our comprehensive, if necessary technical, review [3]. The extension of these results to the cosmological constant version of GR is straightforward; an example is the rather more involved notion of energy there [4]. During this highly productive period, much of it at a distance in those medieval pre-electronic days, we became aware of parallel, if perhaps less comprehensive, results being achieved by Dirac at Cambridge (good thing we were initially unaware of them or we might never have dared enter in such a competition!). Nowadays of course, GR— more specifically, the interplay of Black Holes with Quantum Mechanical problem— is a central preoccupation of EP and QFT theoreticians (for lack of better problems?), as well as the many aspects of the recently observed gravitational waves, whose properties conform perfectly with our predictions. Perhaps our conceptually most surprising initial result came from our (first-order,

"Palatini") reformulation of the original geometrical Einstein-Hilbert action: It came with a vanishing "Hamiltonian", the whole engine of a dynamical system's evolution! Instead, there were four constraints on the six (p,q) pairs per point—p was, as usual, essentially the time derivative of q (here the spatial metric's components). At first floored by this seemingly nonsensical result, we soon realized that this was just the then obscure "Jacobi" form of the action principe, one that can always be achieved by covariantizing any action—except that here the theory came in "already covariantized" form: there are no preferred time or space coordinates to fall back on, unlike in special relativity! This is the central point of any coordinate-invariant system. The Hamiltonian can only be defined, as it should, once a time coordinate is chosen—to measure with respect to which a system's evolution can be charted—no time, no time evolution! Jacobi probably could not envisage this deep physical fact in the early 19th Century. [As a pedagogical remark, I had just taught my first ever course, at Harvard, on classical mechanics, whose boredom made me convert it into one on variational principles in mechanics, whence the Jacobi form. We always learn by teaching.] The first order form was also to play a major role in some of my later work, first enabling me to derive the necessity for, and the action of, full GR from its linear limit (in one line), second to serve as the basis of the derivation, with the late Bruno Zumino, of Supergravity (the gauge unification of, necessarily quantized, GR with a massless spin 3/2 fermion) also in simple analytic form.

## Acknowledgements

This work was supported by the U.S. Department of Energy, Office of Science,